\documentclass[prb,aps,twocolumn,floats,showpacs,amsmath,amssymb]{revtex4}
\usepackage{bm,graphicx}
\newcommand{\be}{\begin{equation}}
\newcommand{\ee}{\end{equation}}
\newcommand{\bea}{\begin{eqnarray}}
\newcommand{\eea}{\end{eqnarray}}
\newcommand{\phrl}[1]{Phys.~Rev.~Lett. {\bf #1}}
\newcommand{\phrb}[1]{Phys.~Rev.~B {\bf #1}}

\newcommand{\bib}{\bibitem}

\begin{document}

\title{A new self-consistent basis generation scheme for polaron and bipolaron systems}

\author{Monodeep Chakraborty}
\email[E-mail: ]{monodeep@postech.ac.kr}
\affiliation{Department of Physics, 
Pohang University of Science and Technology, Pohang, 790-784, Korea}
\author{B. I. Min  }
\email[E-mail: ]{bimin@postech.ac.kr}
\affiliation{Department of Physics, 
Pohang University of Science and Technology, Pohang, 790-784, Korea}

\date{\today}


\begin{abstract}
We have developed a new self-consistent scheme of generating variational basis 
based on the exact-diagonalization, which can be applied efficiently to 
various types of electron-phonon systems.
This scheme is quite general and brings down the size of the variational space 
by an order of magnitude or even more in some cases 
to reproduce the most precise
ground state energies and correlation functions available in the literature. 
This method has enormous potential for application to systems 
with more electrons or in higher dimensions, 
which are still beyond the reach of exact-diagonalization 
just because of the sheer size of their variational space needed to get
reasonably convergent results.
\end{abstract}

\maketitle
\section{Introduction}

The polaron physics in the presence of electron-phonon ({\it e-ph}) and 
electron-electron ({\it e-e}) interactions
is an important subject of interest in the condensed matter physics.\cite{Alex3,mott1}
Enormous amount of analytical and numerical work has been performed 
in an effort to unravel the intriguing polaron-related physics 
in various interesting systems, such as CMR manganites,\cite{cmr1,cmr2}
organic superconductors,\cite{org1} and 
high-T$_c$ superconductors.\cite{sc1,sc2}
Microscopic models employed for the polaron (many-polaron) physics
in the above systems are the Holstein-Hubbard and Fr\"{o}hlich-Hubbard models. 

The analytical approaches to solve the above Hamiltonians are 
mostly based on the many-body perturbation theory 
and so their applicabilities are often
restricted to weak and strong-coupling regimes of the {\it e-ph} coupling.
Accordingly, they are less applicable to the physically interesting crossover regime.
Instead, precise numerical methods are employed, such as
variational approaches based on the exact-diagonalization (VAED), 
the density matrix renormalization group, and the quantum Monte-Carlo scheme.
The VAED is highly accurate for the polarons and bipolarons
in the dilute limit.
The first VAED calculations were reported\cite{Trug3,Trug4}
more than a decade ago and
they were quite accurate for large polarons and more so in the physically
interesting crossover regime. 
A very rudimentary effort to increase the scope of
the VAED method to the strong coupling regime and to the crossover regime for the
adiabatic polarons was made by Chakrabarti {\it et al.},\cite{Atis1}
who started with two initial states
(the zero phonon state and the state with a large number of phonons at the 
electron site) to meet with some success. 
More recently, the Lang-Firsov (LF) idea has been incorporated
in the variational scheme,\cite{Alt2,Mono2} 
which makes the method more precise through out the parameter
regimes at least for polaron and bipolaron in the one-dimension (1D). 
The scheme of Alvermann {\it et al.},\cite{Alt1} in which a shifted oscillator
state (SOS) is considered over the traditional VAED states,
is very precise to account for the most
difficult adiabatic polarons in the crossover regimes. 

The VAED method has been highly successful in the dilute limit, 
but is applicable to only one or two particle system.
Real systems, however, require the study of {\it e-ph} models 
with more than two electrons.\cite{hohenalder1,Fehske2}
The question we have addressed in this paper is whether there is further scope to improve
the VAED method that could study more than two electron systems.
To this end, we have developed a new scheme, the self-consistent VAED (SC-VAED) method,
which is quite efficient and general.
In the SC-VAED,
instead of generating the variational basis in a single step
as done in traditional VAED method, we start with a small basis to calculate the ground state,
and then restart the whole process only with a few initial states, which 
carry the significant probabilities of the ground state wave function. 
This process is repeated till the desired accuracy is achieved. 

This paper is organized as follows.
In section II, we introduce the Hamiltonian in its most general form,
which incorporates the {\it e-e} and {\it  e-ph} interactions,
within the Holstein-Hubbard and Fr\"{o}hlich-Hubbard model.
In section III, we describe the basis generation scheme in the SC-VAED method.
In section IV, we compare the ground-state energies of different Holstein and Fr\"{o}hlich systems 
obtained by using the SC-VAED with those available in the literature.
We also discuss the electron-lattice correlation function for a large polaron
and the bipolaron mass in the strong {\it  e-ph} coupling regime
to highlight the applicability of the SC-VAED method to different regimes of {\it e-e}
and {\it  e-ph} interactions.
Conclusion follows in Section V.

\section{The Hamiltonian}

                The general Hamiltonian on a discrete  lattice,\cite{Fehske1,Alex1}
which includes both the {\it e-e} and {\it e-ph} interactions, 
is considered :
\bea
H = &&- \sum_{i,\sigma}(t c_{i,\sigma}^{\dag} c_{i+1,\sigma} + h.c)
+ \omega \sum_i a_i^{\dag} a_i \nonumber \\
&&+ g\omega \sum_{i,j,\sigma}f_{j}(i) n_{i,\sigma} (a_{i+j}^{\dag}
+ a_{i+j}) \nonumber \\
&&+ U\sum_{i}n_{i,\uparrow}n_{i,\downarrow},
\eea
                              where $c_{i,\sigma}^{\dag}$($c_{i,\sigma}$)
creates (annihilates) an electron of spin $\sigma$,
and $a_{i}^{\dag}$ ($a_{i}$) creates (annihilates) a
phonon at site $i$. The third term represents the coupling of
an electron at site $i$ with an ion at site $j$, where $g$ is the
dimensionless {\it e-ph} coupling constant. 
$f_{j}(i)$ is the long-range {\it e-ph} interaction, 
the actual form of which is  given by\cite{Fehske1}
\bea
f_{j}(i) = \frac{1}{(|i-j|^{2} +1 )^{\frac{3}{2}}}.
\eea
$U$ is the on-site Hubbard {\it e-e} interaction strength. 
 We set the electron hopping $t$=$1$ for the numerical calculations
and all energy parameters are expressed in units of $t$.

The Holstein model is recovered by setting $i$=$j$  in Eq. 2. Incorporating the
whole Fr\"{o}hlich interaction is numerically impossible. 
Bonca and Trugman\cite{Trug2} simplified this model by placing ions
in the interstitial sites located between Wannier orbitals,
and then considered just the nearest-neighbor {\it e-ph} 
interaction (F2H model), \cite{Mono2}
which corresponds to the case of $f_{i \pm \frac{1}{2}}(i)$=$1$ and zero otherwise. 
This case has been discussed in detail by Bonca and Trugman\cite{Trug2} 
and Chakraborty {\it et al.}.\cite {Mono2}
Chakraborty {\it et al.}\cite {Mono2} also investigated
the effect of extending the spatial extent of {\it e-ph} interaction
(F3H and F5H models).
In the presence of $f_{j}(i)$ interaction,
the {\it e-ph} coupling constant $\lambda$ is 
defined by\cite{Fehske1,Trug2} 
\bea
\lambda= \frac{\omega g^{2}\sum_{l}f_{l}^{2}(0)}{2t}.
\eea

\begin{table*}[t]
\caption {
The ground state energies E$_{0}$'s for different {\it e-ph} systems
obtained by the present SC-VAED are compared with the most precise E$_{0}$'s obtained by the VAED in the literature. 
The basis sizes $N_{Basis}$ used to obtain E$_{0}$'s are provided together.
D, Model, and N$_\mathrm{e}$ represent the dimension, the Hamiltonian (Holstein (H) 
or Fr\"{o}hlich-2 (F2)), and the number of electrons in the system, respectively. 
$\omega$, $\lambda$, {\it U} denote phonon frequency, {\it e-ph} coupling, Coulomb interaction,
respectively, in units of $t$. 
}
\begin{ruledtabular}
\begin{tabular}{c l c l | l l c | l l | l l l} 
 Case & D & Model & N$_\mathrm{e}$ &  $\omega$ & $\lambda$ & {\it U} &  E$_{0}$(SC-VAED) & $N_{Basis}$ &  E$_{0}$(VAED) & $N_{Basis}$ & Literature \\  \hline   
 1&  1D& H  &   1  &  1& 0.5& 0 & -2.46968472393  &2.4$\times 10^4$  & -2.469684723933  &8.8$\times 10^4$ &  Ref.[\cite{Trug3,Trug4,Atis1}] \\ 
 2&  1D& H  &   1  &0.1& 1.0& 0 & -2.53800667     &5.0$\times 10^5$  & -2.53800669      &3.0$\times 10^6$ &  Ref.[\cite{Alt1,Mono2}] \\   
 3&  2D& H  &   1  &  2& 0.5& 0 & -4.81473577884  &5.0$\times 10^5$  & -4.814735778337  &5.5$\times 10^6$ &  Ref.[\cite{Trug4}] \\   
 4&  3D& H  &   1  &  3& 0.5& 0 & -7.1623948637   &1.9$\times 10^5$  & -7.1623948409    &7.0$\times 10^6$ &  Ref.[\cite{Trug2,Mono2}] \\   
 5&  1D& H  &   2  &  1& 0.5& 0 & -5.4246528      &1.4$\times 10^5$  & -5.4246528       &2.2$\times 10^6$ &  Ref.[\cite{Mono2}] \\   
 6&  1D& H  &   2  &  1& 2.0& 0 & -16.25869250598 &2.0$\times 10^5$  & -16.25869250598  &1.7$\times 10^7$ &  Ref.[\cite{Trug2,Mono2}] \\   
 7&  1D& F2H&   2  &  1& 0.5& 1 & -5.82261974     &2.75$\times 10^5$ & -5.822621        &3.0$\times 10^6$ &  Ref.[\cite{Mono2}] \\   
\end{tabular}
\end{ruledtabular}
\end{table*}

\section{The SC-VAED}

Figure 1 provides a schematic picture of generating the basis state in the VAED.
Starting from the initial state with two electrons and zero phonon,
new translationally invariant states are generated by a single operation 
of the Holstein Hamiltonian on the initial state.
As mentioned above, 
the VAED method is restricted to one or two particle system. 
We have thus tried to improve the VAED method 
to deal with systems with more than two electrons.

\begin{figure}[t]
\includegraphics[scale=0.30]{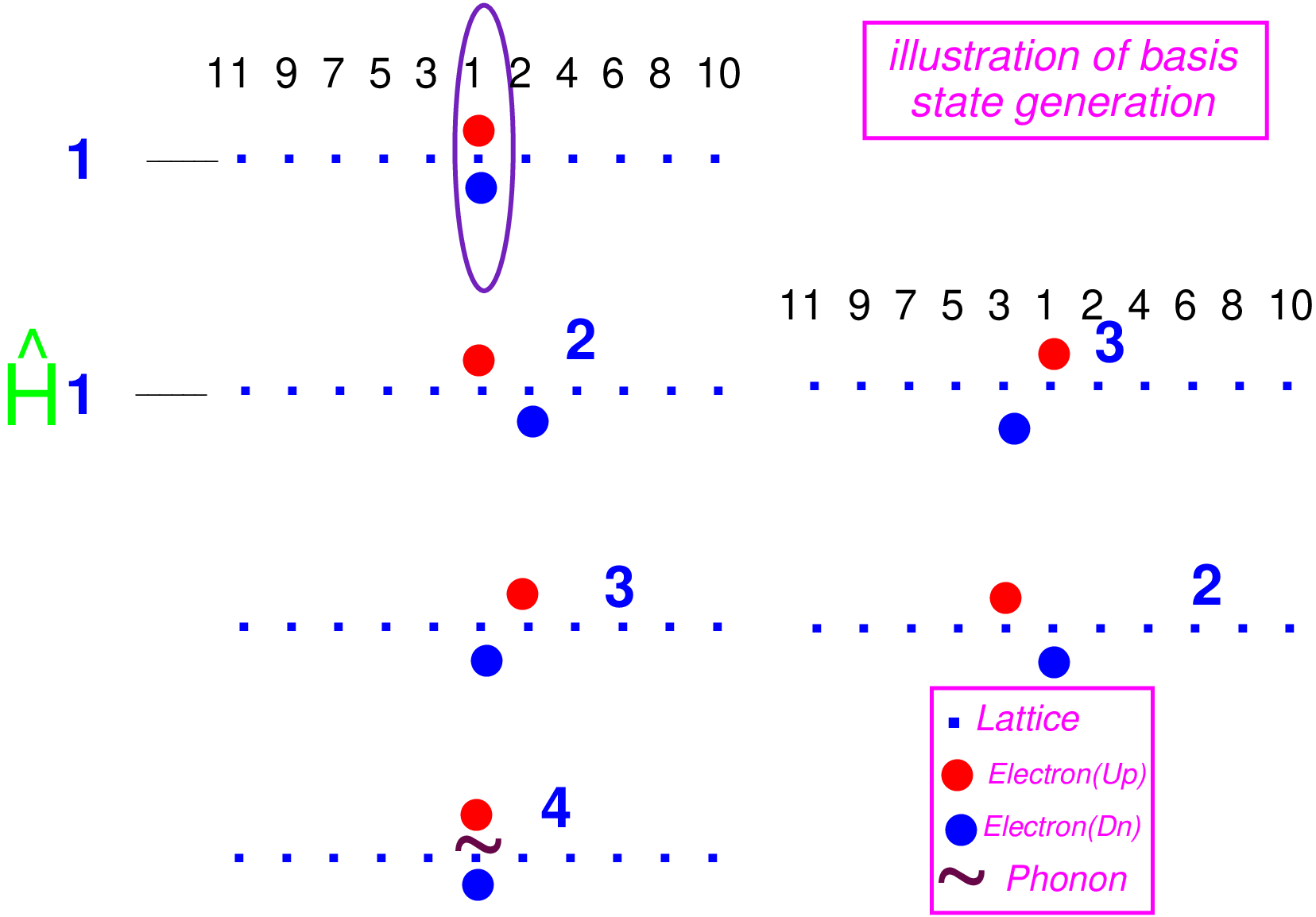}
\caption{\label{f1} (Color online)
The illustration of basis state generation from the initial singlet Holstein bipolaron state.
Two electrons with spin-up (red ball) and spin-down (blue ball) are located 
at the lattice site $1$.
New sates are generated by the single operation of off-diagonal term of the Holstein Hamiltonian. 
If two states are related by translational symmetry, then a single state is retained.\cite{Trug3,Trug1,Mono2}
The tilde mark represents the phonon.
 }
\end{figure}

\begin{figure}[b]
\includegraphics[scale=0.30]{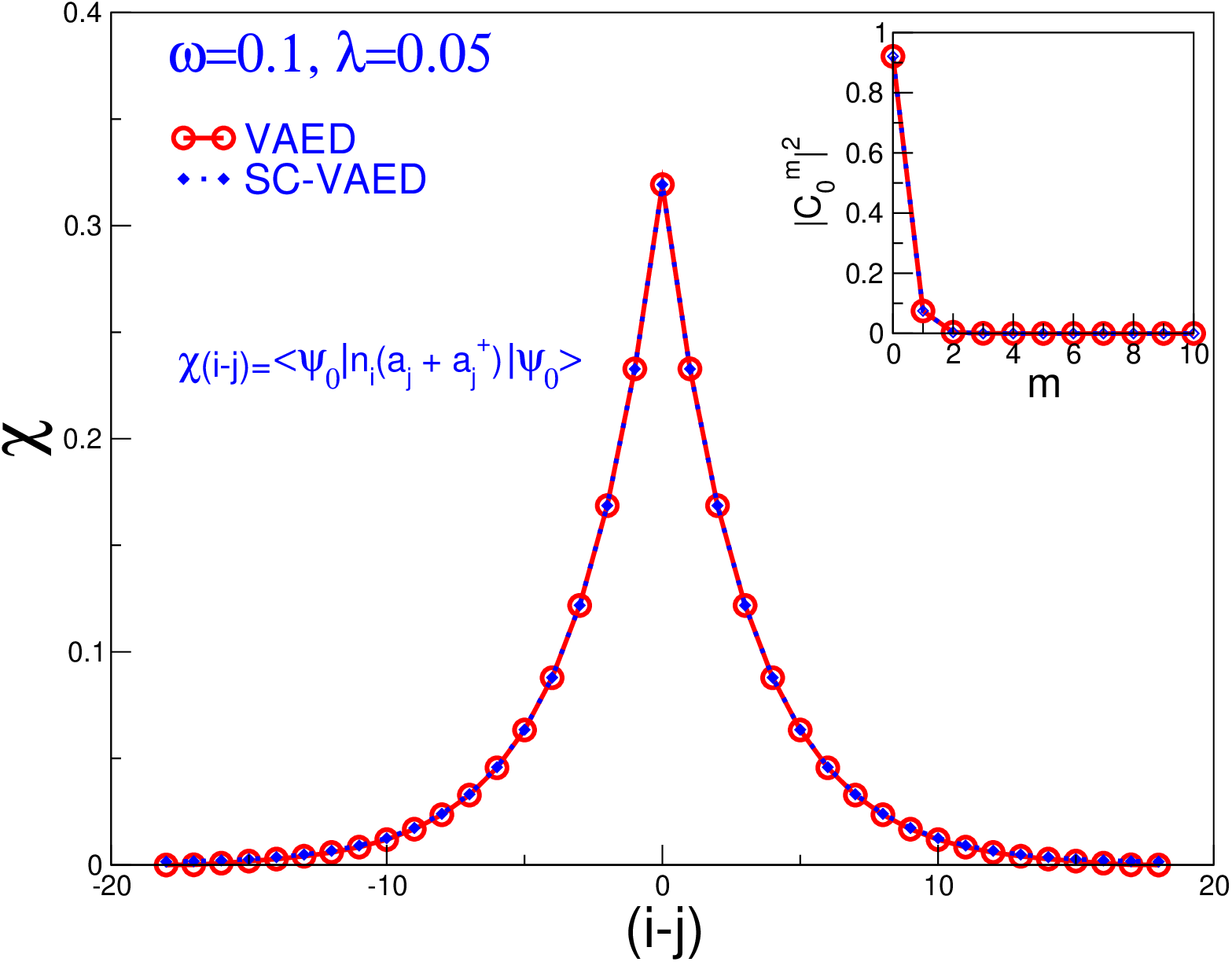}
\caption{\label{f2} (Color online)
The electron-lattice correlation function $\chi$(i-j) for
a large polaron at $\omega$=$0.1$ and $\lambda$=$0.05$. The inset shows the
weight of $m$-phonon states for the ground state polaron.\cite{Fehske1} 
We compare the quantities calculated by using the SC-VAED and the VAED method. 
Here the size of the basis used in the SC-VAED is $26000$, 
whereas the VAED requires much larger basis of $731027$ states.\cite{Trug2,Atis1}
}
\end{figure}

We have made systematic analyses of the ground state wave-functions
of already well-studied systems, and found that most of the probability 
of the wave-function is contained in a few number of states. 
On the basis of this finding, we devise a scheme that throws away not so
important states and builds upon the higher weighted states.
Namely, for a given lattice size, 
instead of generating the variational space at once,
we first generate small number of states (say $10000$) and obtain 
the ground state wave-function and energy. 
We pick up a few of the states with the highest probability (say $1000$). 
Now a basis of bigger size than the first basis (say $12000$) is
generated with these (say $1000$) states as the starting states. 
We repeat this process with increasing the size of the basis at each step. 

The result is quite encouraging.
As shown below, this scheme reproduces the best available results in all parameter regimes with a basis much
smaller than used before. The higher phonon number states are picked up by the
self-consistency cycles. We check the convergence by comparing the converged
energies for different lattice sizes.  

\section{Results}

The notable  feature of our development is that we are in a position 
to reproduce the benchmark results at much lesser computational cost. 
Table 1 shows the ground state energies 
for different {\it e-ph} systems obtained by the SC-VAED, which are compared with
the best results available from literature. 
The strength of the traditional VAED exists
for small {\it e-ph} coupling and the intermediate phonon regime 
(case 1 in Table 1).\cite{Trug1,Trug2,Trug3,Trug4} 
We are able to obtain similar precision in the SC-VAED with a basis size much smaller. 
The VAED fails to maintain its high standard for the adiabatic case 
with intermediate {\it e-ph} coupling. The SOS-VAED scheme of Alvermann {\it et al.}\cite{Alt1} 
is an excellent approach to overcome this limitation of the VAED 
(case 2 in Table 1). 
Incorporation of the LF idea\cite{Mono2} also yields
similar success, but with a much bigger basis size. 
Noteworthy is that the SC-VAED scheme obtains 
the same precision in this regime too, again with a smaller basis size.  
Chakraborty {\it et al.}\cite{Mono2} showed that
the strong coupling regime could be handled efficiently 
with the LF-VAED (case 6 in Table 1).
The SC-VAED describes two-electron Holstein-Hubbard bipolaron system 
as efficient as the LF-VAED but at a much lower computational cost. 
The SC-VAED works equally well for the Fr\"{o}hlich system too (case 7 in Table 1).

\begin{figure}[b]
\vskip 0.5cm
\includegraphics[scale=0.30]{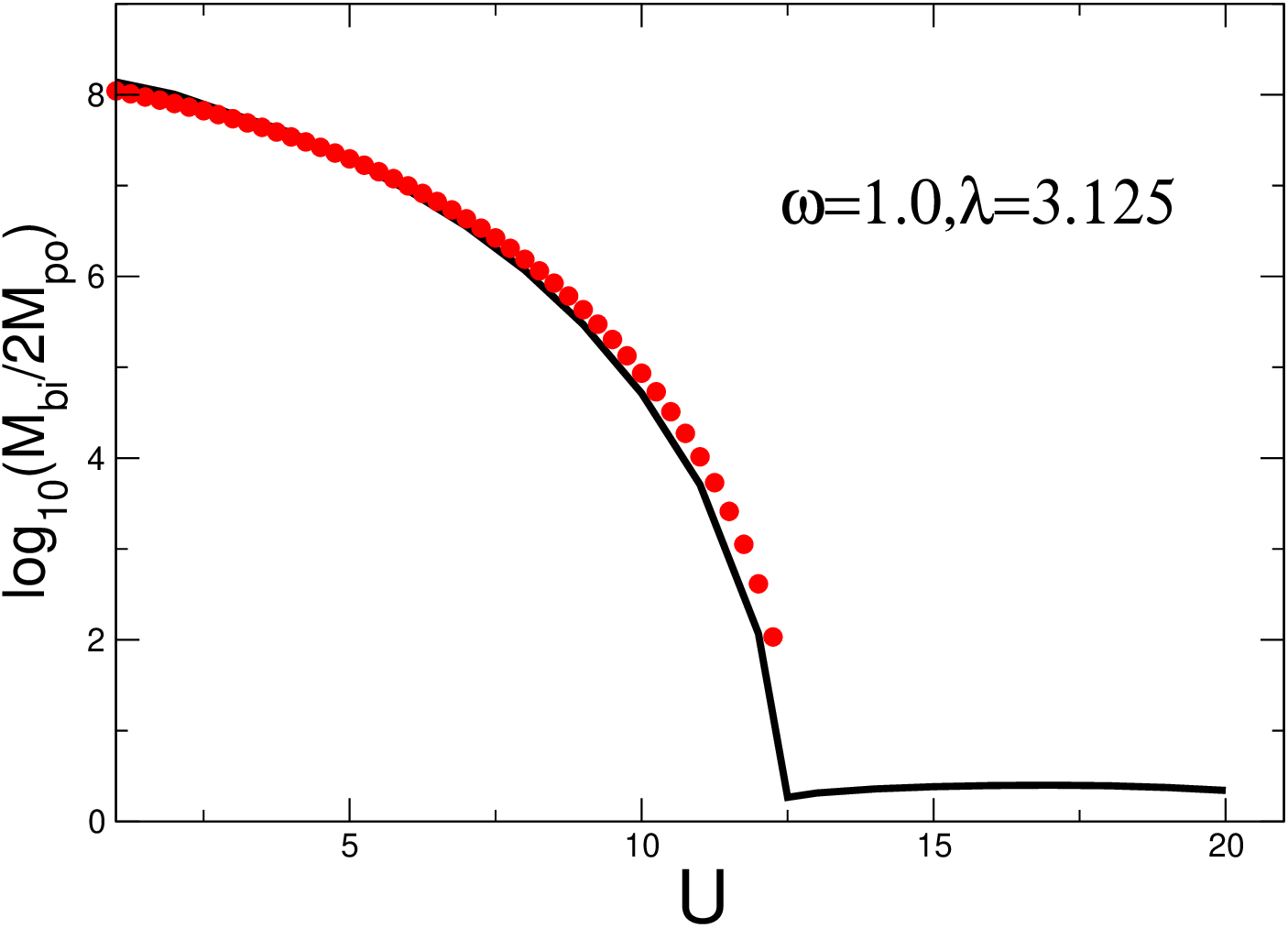}
\caption{\label{f3} (Color online)
Effective mass of a Holstein bipolaron as a function of $U$ 
at $\omega$=$1.0$ and $\lambda$=$3.25$,
which is normalized by twice the mass of polaron at the same parametric regime.
The SC-VAED results (solid line) are compared with analytic results (dotted line) obtained from 
the second order strong coupling perturbation theory.\cite{Alex1,Trug1,Mono2}
}
\end{figure}

The comparison in Table 1 clearly manifests that the SC-VAED scheme
indeed brings down the numerical burden and thus extends 
the ambit of the method to more difficult parametric regimes 
and to more number of  particles.
The price that one has to pay for this method is 
to make the self-consistent basis at each parameter of the calculation. 
But this is a small price to pay in view of its advantages.


Now we consider two different systems
in different regimes to explain the utility of our development. 
Let us first consider a typical large polaron system. 
Figure 2  shows the static electron-lattice correlation function \cite{Trug3,Atis1,Mono1} 
in the adiabatic regime ($\omega$=$0.1$)
and at  very low {\it e-ph} coupling ($\lambda$=$0.05$). 
The inset shows $|C_{0}^{m}|^{2}$, which corresponds to
the weight of the phonon states as defined by Fehske {\it et al.}.\cite{Fehske1} 
The SC-VAED results match excellently with the VAED results.\cite{Trug3,Atis1,Mono1}
The VAED results were calculated with a basis size of $731027$, whereas the SC-VAED
calculations were done with a basis size of $26000$. 
Although the lattices sizes are similar, 
the self-consistent cycles get rid of the higher phonon number
states that do not contribute to the ground state wave-function significantly,
thus keeping intact the accuracy with a much smaller basis.

We next consider the case of extremely strong {\it e-ph} coupling.
Figure 3 shows the effective mass of a Holstein bipolaron 
as a function of on-site {\it e-e} Hubbard interaction $U$ 
at $\omega$=$1.0$ and $\lambda$=$3.25$.\cite{Trug1,Mono2}
It is normalized by twice the mass of the polaron at that parameter regime.
It is seen that the SC-VAED result is in close agreement with the
analytical calculation.\cite{Alex2,Trug1,Mono2} 
It should be noted that no prior numerical calculation
has been attempted at this regime for bipolarons.

The above two examples demonstrates the potential applicability of the SC-VAED scheme 
to any {\it e-ph} coupling regime 
and to different polaron and bipolaron systems
of both Holstein and Fr\"{o}hlich varieties.

\section{Conclusions}

We have developed the self-consistent variational approach (SC-VAED),
which not only reproduces the most precise results with a much lesser computational effort
but also increases the scope of variational approach to much bigger systems. 
The SC-VAED method is simple and easily implementable. 
The real benefit of the SC-VAED scheme will become evident
when applied to problems involving more electrons in higher dimension, 
suggesting that the SC-VAED is a very promising method with a lot of applicability. 

\acknowledgements
This work was supported by the NRF (No.2009- 0079947) and the
POSTECH Physics BK21 fund. 
Stimulating discussions with H. Fehske and A. Alvermann are
gratefully acknowledged.

\end{document}